\documentclass[a4paper]{llncs}

\usepackage{amssymb}
\usepackage{graphicx}
\usepackage{subfigure} 

\usepackage{comment} 

\usepackage{url}

\let\llncssubparagraph\subparagraph
\let\subparagraph\paragraph
\usepackage[compact]{titlesec}
\let\subparagraph\llncssubparagraph

\newcommand{\keywords}[1]{\par\addvspace\baselineskip
\noindent\keywordname\enspace\ignorespaces#1}

\begin{document}

\mainmatter  

\title{Weighted Shortest Path Models: A Revisit to the Simulation of Internet Routing}


%
%
\author{Mingming Chen\inst{1} \and Jichang Zhao\inst{2,3} \and Xiao Liang\inst{3} \and Ke Xu\inst{3}}
%

\institute{Department of Computer Science, Rensselaer Polytechnic Institute
\and School of Economics and Management, Beihang University
\and State Key Laboratory of Software Development Environment, Beihang University
\\ chenm8@rpi.edu, jichang@buaa.edu.cn, \{liangxiao, kexu\}@nlsde.buaa.edu.cn
}

%
%

\maketitle

\begin{abstract}
Understanding how packets are routed in Internet is significantly important to Internet measurement and modeling. The conventional solution for route simulation is based on the assumption of unweighted shortest path. However, it has been found and widely accepted that a packet in Internet is usually not transmitted along the unweighted shortest path between its source and destination. To better simulate the routing behavior of a packet, we thoroughly explore the real-world Internet routes and present a novel local information based simulation model, with a tuning parameter, which assigns weights to links based on local information and then simulates the Internet route with weighted shortest path. Comparisons with baseline approaches show its capability in well replicating the route length distribution and other structural properties of the Internet topology. Meanwhile, the optimal parameter of this model locates in the range of $(0,2)$, which implies that a packet inside the Internet inclines to move to nodes with small degrees. This behavior actually reflects the design philosophy of Internet routing, balancing between network efficiency and traffic congestion.

\end{abstract}

\section{Introduction}
It is very important to track how a packet is routed inside Internet, which is of great help to Internet measurement and modeling. Specifically, it could assist the deployment of sources and destinations for traceroute which is a generally employed tool to sample Internet \cite{Skitter,Scamper,DIMES,iPlane,Heuristics,Rocketfuel,NTC,TraceroutePattern}. However, the realistic routing process in Internet is complicated, which makes the simulation of the Internet routes be a significant challenge. For simplicity, almost all previous approaches assumed that a route in Internet is just an unweighted shortest
path between the source and the destination \cite{TheoryAndSimulations,SamplingBiases,AccuracyScalingINETMap,RelevanceOfMassivelyQualitative,BiasTracerouteSampling}.
Based on this assumption, \cite{TheoryAndSimulations}
proposed three mechanisms for the exploration process of traceroute:
USP (Unique Shortest Path), RSP (Random Shortest Path), and ASP (All
Shortest Path). While USP model, also called
\textit{(k,m)-traceroute}, is the most widely used one. Namely, the
classical approach to simulate the Internet routing is to consider it as
an unweighted shortest path model (\textit{USPM}). However, generally, real routes do not have the same properties as the unweighted
shortest path model assumes, which has already been pointed out in
\cite{EndToEndRouting,SimulatingInternetRoute,RealSizeOfSampledNetwork} and could be learned from their route length
distribution difference in Subsection~\ref{subsec:motivation}. Instead, the real routes are impacted by many factors, including commercial agreements, traffic congestion, and administrative routing policies \cite{RouingPolicyImpact}.

To improve the route simulation, recently a node degree model (\textit{NDM}) \cite{SimulatingInternetRoute} was proposed, whose idea is as follows. Two paths are calculated initially, one starting from the source and the other from the destination, where the next selected node on the path is always the highest degree neighbor of the current node. The computation terminates when it reaches a situation where a node is the highest degree neighbor of its own highest degree neighbor. Then, one of two cases applies: either the two paths have met at a node, or they have not. In the first case, the route produced by the model is the discovered path. In the second case, the model tries to find a shortest path between the two paths, and then obtains the route by merging the two paths and the shortest path. However, there is still a large gap between this model and the real Internet routing. It is still somewhat based on the unweighted shortest path method and has no physical senses for its omitting the
principles of link dynamic \cite{PathDiversity}, traffic dynamic \cite{TrafficDynamicLocal}, etc. For
instance, it does not take the path attributes, such as delay, loss
rate, and available bandwidth, into account. Also, it always routes
to the large degree nodes, that is to say, routes to the core of the
network, without considering the traffic load balance \cite{LoadBalance,LoadBalance2}. Imagining that
there is a link between two nodes at the border of Internet, it is unnecessary
and unreasonable for one of them via the backbone of the
network in order to reach the other.

Thus, a new and more accurate model with physical meanings is still needed. In this paper, we thoroughly investigate the real-world
traceroute traces and propose two new simulation models, \textit{Local Information Based Model} (\textit{LIM}) and \textit{Path Feature Based Model} (\textit{PFM}) both with tunable parameters $\alpha$, which weigh links flexibly and then simulate the Internet routing with weighted shortest path. The weights to links that our models assign are intent to mimic the link dynamic \cite{PathDiversity}, traffic dynamic \cite{TrafficDynamicLocal}, etc. The comparison of our two models with \textit{USPM} and \textit{NDM} shows that our two models, especially \textit{LIM}, could more accurately simulate the real Internet routes. In addition, the optimal parameter of \textit{LIM} locates in the range of $(0,2)$, which implies that a packet in Internet tends to move to small degree nodes. This behavior reflects the trade-off between network efficiency and traffic load balance in the design philosophy of Internet routing.

The rest of this paper is organized as follows. First, in
Section~\ref{sec:preliminaries} we provide the preliminaries. Then
we present the motivation of this work and give an introduction to
our new models in Section~\ref{sec:models}.
Section~\ref{sec:evaluation} evaluates our models and analyzes the
underlying mechanisms for optimal parameters.
Finally, we conclude this work briefly in
Section~\ref{sec:conclusion}.

\section{Preliminaries}
\label{sec:preliminaries}

The mapping of the topological structure is greatly important
for a better understanding of Internet. Current explorations
still rely on the extensive use of traceroute: one
collects routes from a limited set of sources to a large set of
destinations, and then merges the obtained paths into a graph. This
method has been adopted by many influential network topology discovery
projects
\cite{Skitter,Scamper,DIMES,iPlane,Heuristics,Rocketfuel,NTC}.
In order to better simulate the Internet routes, we deeply study the real
traceroute traces of two network topology datasets: iPlane
\cite{iPlane} and skitter \cite{Skitter}. The routes of both datasets are undirected and unweighted. Also, we evaluate
our two models and compare them with other models on these two datasets.

\textbf{iPlane:} The Information Plane performs traceroutes from about two hundreds vantage points every day to map the worldwide IPv4 network topology, and uses the structural information to predict path properties between arbitrary end-hosts. The dataset we use was collected on June, 30, 2011. However, the set of destinations varies among the sources. Therefore, for the convenience of comparison, we obtain a set of 140 sources to a common set of 20,419 destinations. The resulting graph has 266,317 nodes and 1,663,170 edges.

\textbf{skitter:} skitter has deployed more than 30 vantage points around the world to map the IPv4 network topology. The dataset we use was collected on October, 15, 2005. Yet, the set of destinations varies among the sources. Similarly, we obtain a set of 21 sources to a common set of 157,290 destinations. The final graph has 771,312 nodes and 1,785,922 edges.

The Internet topology can be naturally modeled as an undirected graph $G=(V,E)$,
where $V$ denotes the set of nodes and $E$ is the set of edges. The number of links of a node is defined as its {\it degree}. Then, the {\it average degree} of a network can be defined as $\langle k \rangle=\frac{2|E|}{|V|}$.
The {\it degree distribution} of a graph is denoted as $p(k)$. For Internet, $p(k)$ follows a power-law distribution $p(k)\sim
k^{-\gamma}$ \cite{PowerLawRelationship}.
In such case, the exponent $\gamma$ could be considered as
an indicator of how heterogeneous this distribution is.
{\it Heterogeneity} of a
network, defined as $H=\frac{\langle k^2 \rangle}{\langle k \rangle^2}$,
is usually used to characterize the nonuniformity of degrees. The {\it clustering coefficient} of a graph is the probability that two nodes are connected, given that they are both linked to a same third node $C=\frac{3N_\bigtriangledown}{N_\vee}$,
where $N_\bigtriangledown$ denotes the number of triangles (set of three nodes with three edges) and $N_\vee$ is the number of connected triples (set of three nodes with at least two edges) in the graph.

\section{New Simulation Models}
\label{sec:models}

In this section, we first describe the motivation that propels us to propose better simulating models for the realistic Internet routing. Then, we give an introduction to our two weighted based shortest path models. 

\subsection{Motivation}
\label{subsec:motivation}
In this subsection, we deeply analyze the real traceroute traces of the two real datasets and the corresponding unweighted shortest paths from all the sources to all the destinations. Unfortunately, there are large differences in the routes length distribution and the node degrees visited probabilities distribution between the real traces and the corresponding unweighted shortest paths.

\subsubsection{Routes Length Distribution}
\begin{figure}[!t]
\centering
\subfigure[iPlane]{
\label{figure2:subfig:a}
\includegraphics[scale=0.265]{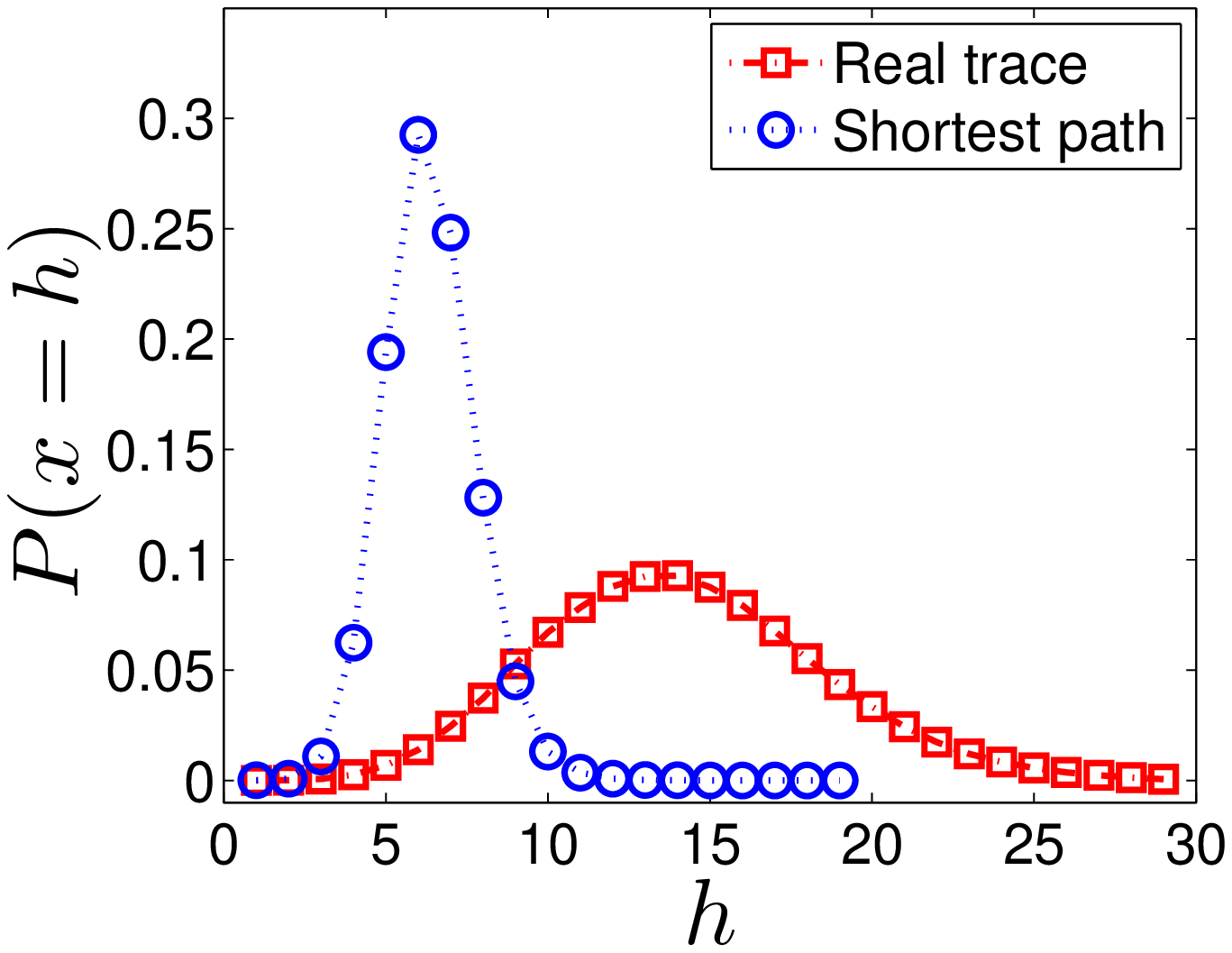}
}
\hspace{2em}
\subfigure[skitter]{
\label{figure2:subfig:b}
\includegraphics[scale=0.265]{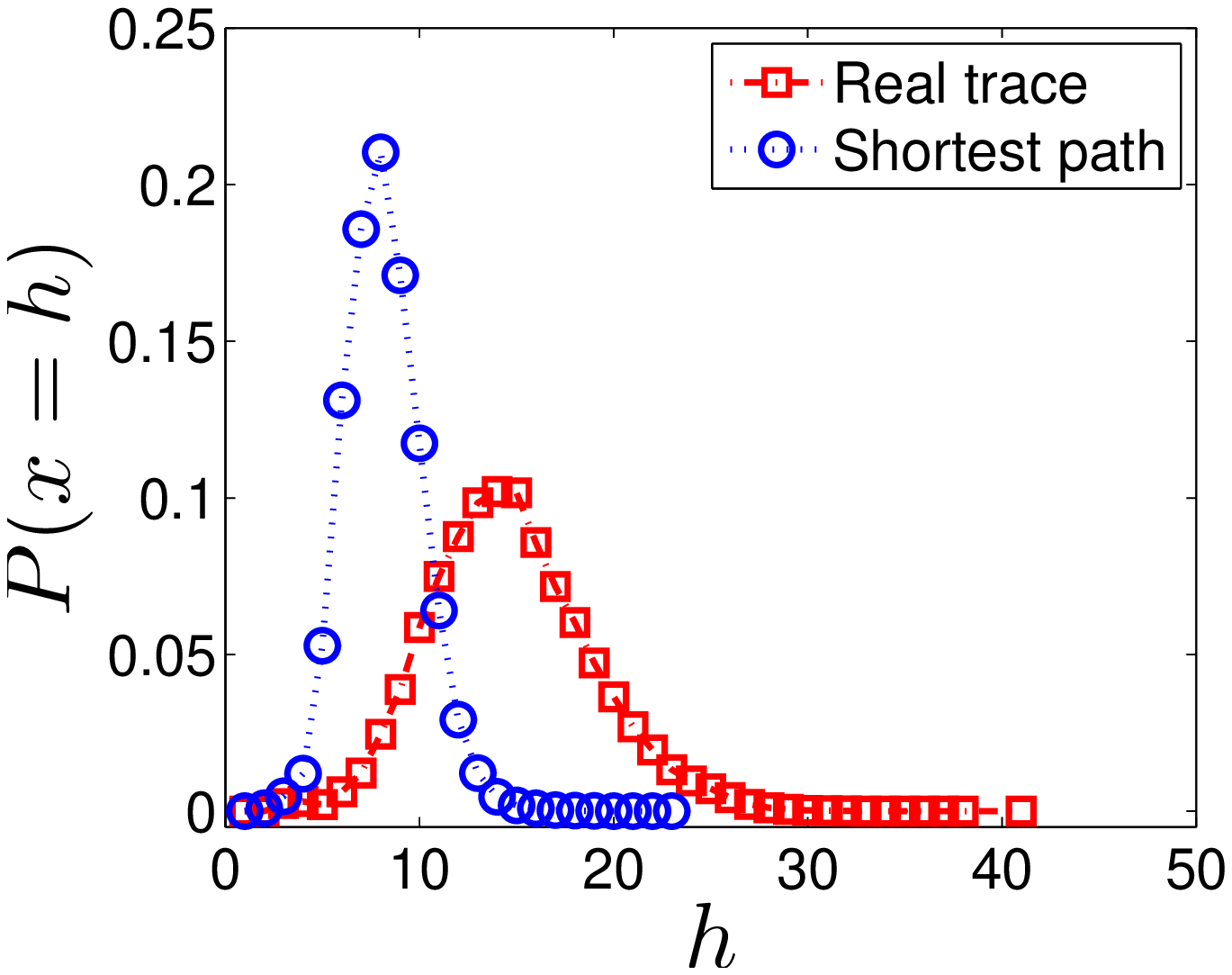}
}
\vspace{-1.7em}
\caption{The routes length distribution of the real traceroute traces and the corresponding unweighted shortest paths of the two real datasets: iPlane and skitter, respectively. Here, the hop of routes is denoted as $h$.}
\label{figure2}
\vspace{-0.8em}
\end{figure}

Fig. \ref{figure2} shows the routes length distributions of the real traces and the corresponding unweighted shortest paths of the two datasets, respectively. As can be seen, the two distributions have large difference with each other. Also, the real traceroute traces have a mean route length of 14.11 hops and 14.57 hops, while the unweighted shortest paths have an average route length of 6.36 hops and 7.39 hops on iPlane and skitter datasets, respectively. The former is almost two times longer than the latter. It strongly indicates that \textit{USPM} is an improper method to simulate the real Internet routes.

\begin{figure}[!t]
\centering
\subfigure[Real traces]{
\label{figure3:subfig:a}
\includegraphics[scale=0.265]{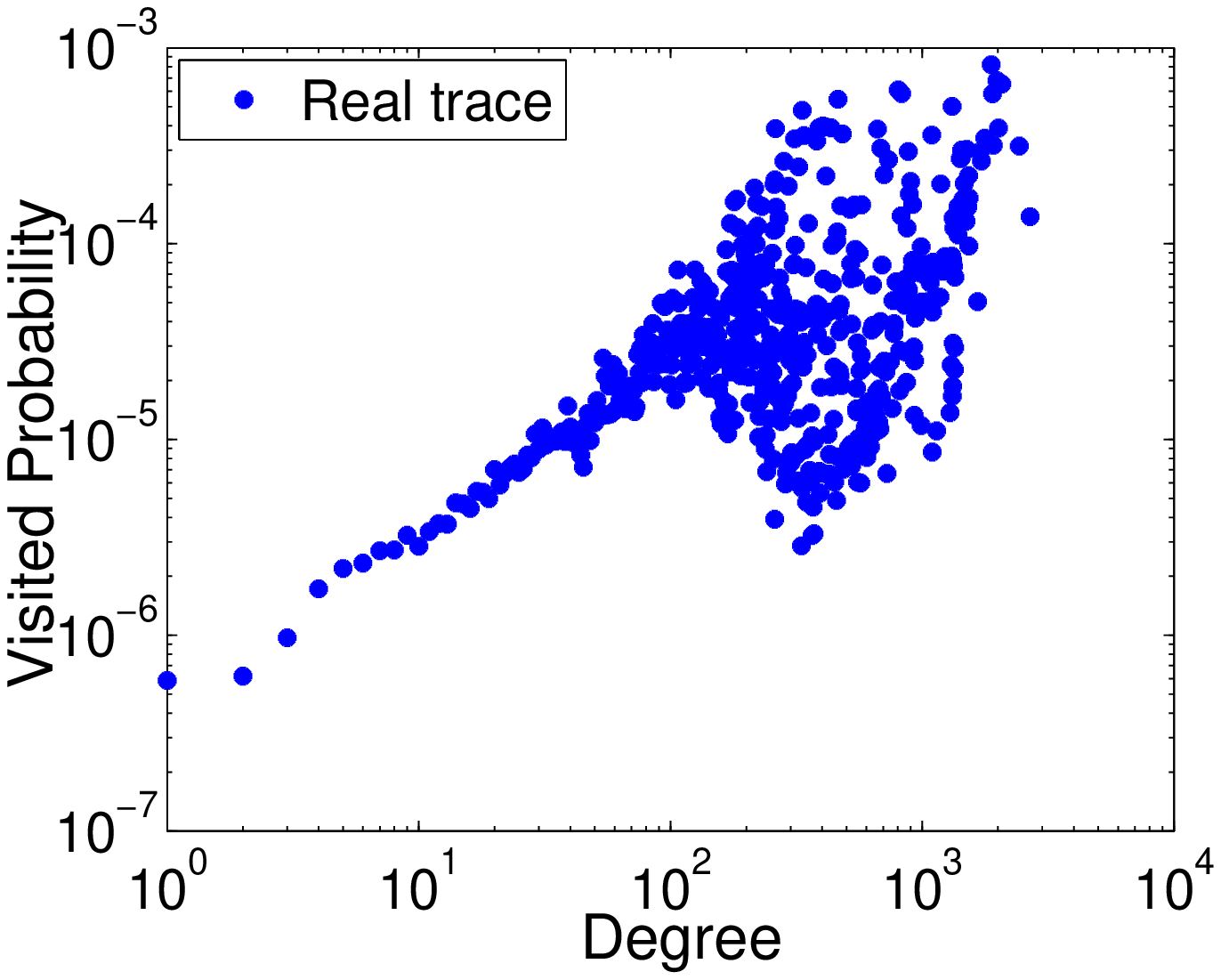}
}
\hspace{2em}
\subfigure[Shortest paths]{
\label{figure3:subfig:b}
\includegraphics[scale=0.265]{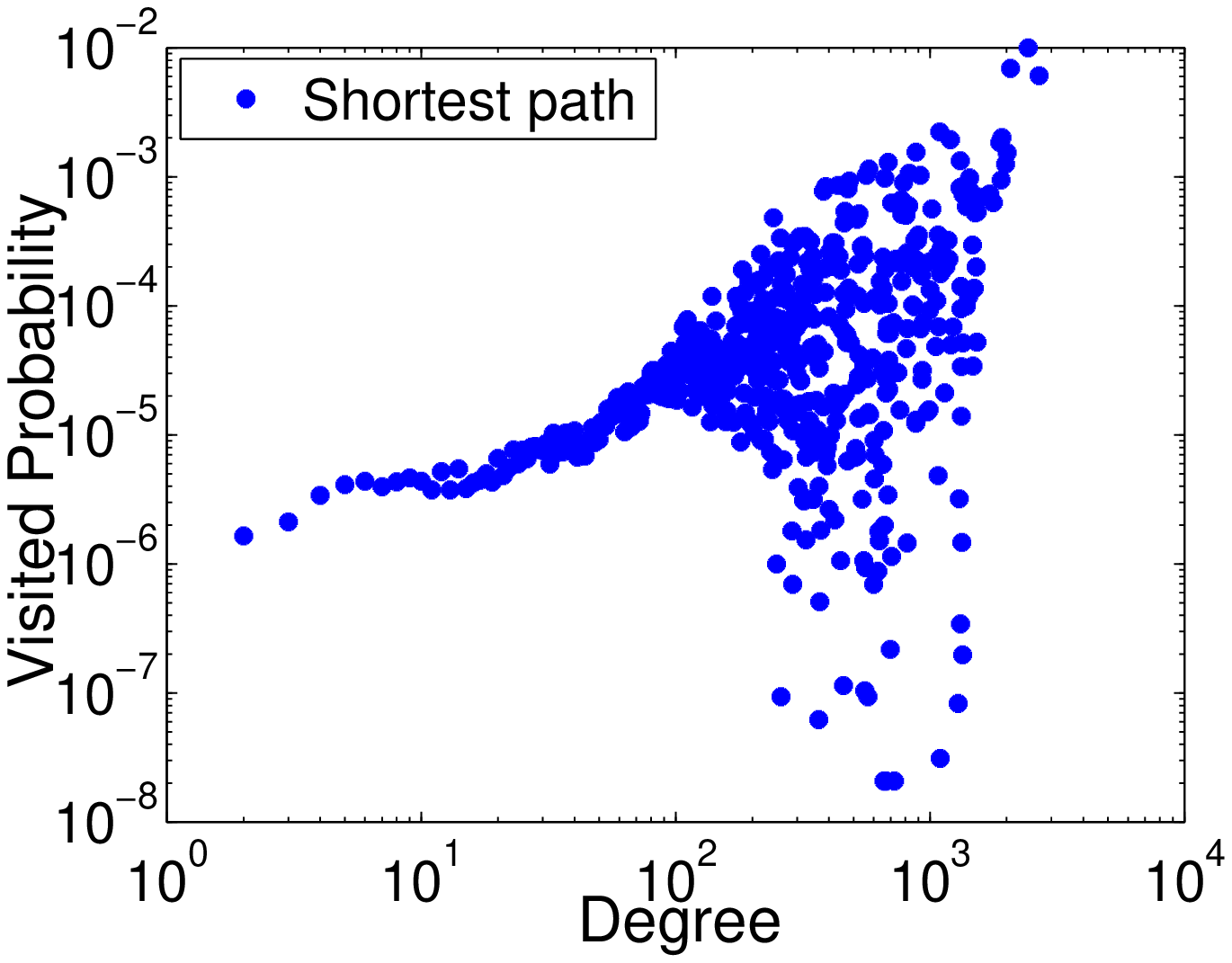}
}
\vspace{-1.7em}
\caption{The node degrees visited probabilities distributions of the real traces and the corresponding unweighted shortest paths on iPlane dataset.}
\label{figure3}
\vspace{-0.8em}
\end{figure}

\begin{figure}[!t]
\centering
\subfigure[Real traces]{
\label{figure4:subfig:a}
\includegraphics[scale=0.265]{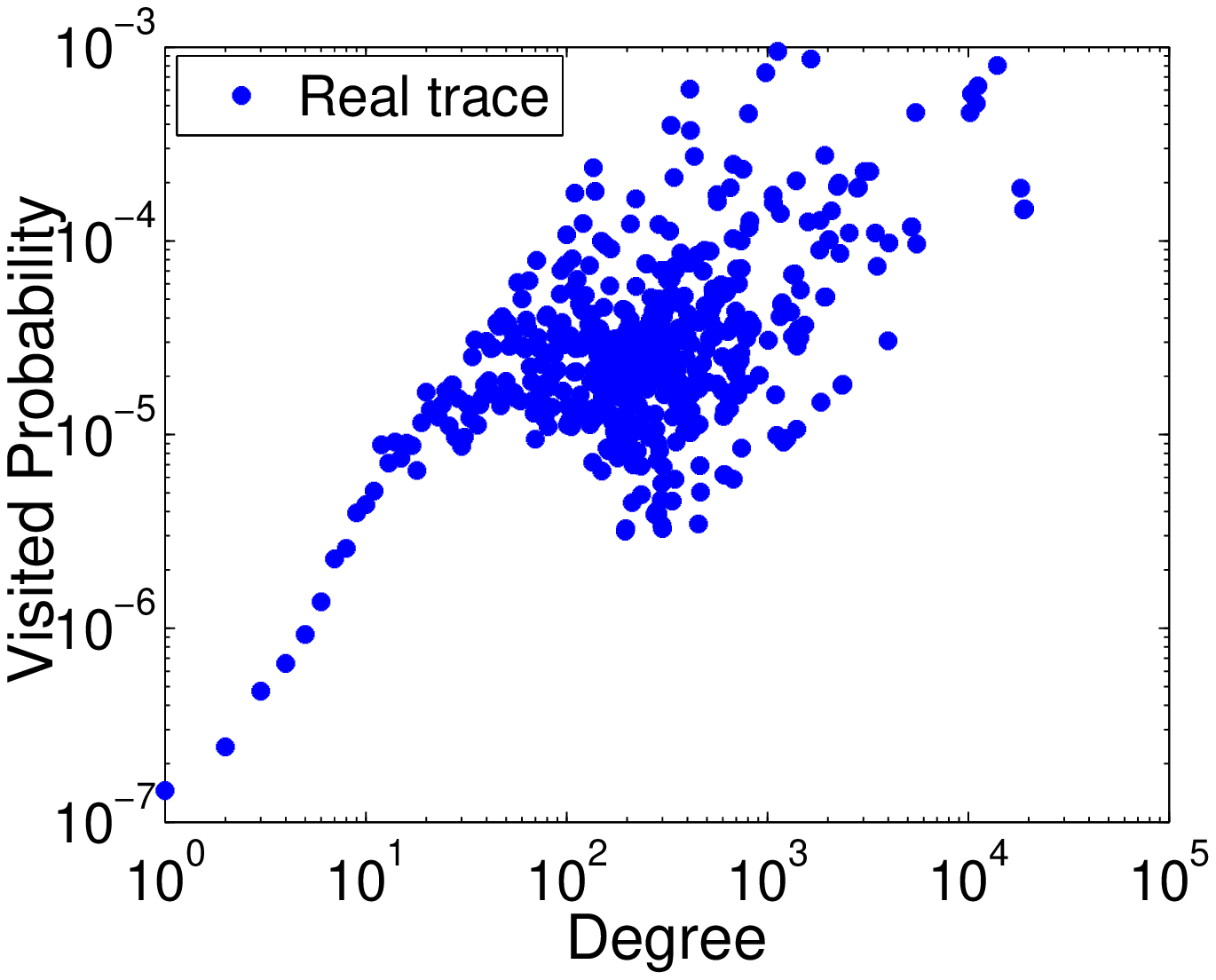}
}
\hspace{2em}
\subfigure[Shortest paths]{
\label{figure4:subfig:b}
\includegraphics[scale=0.265]{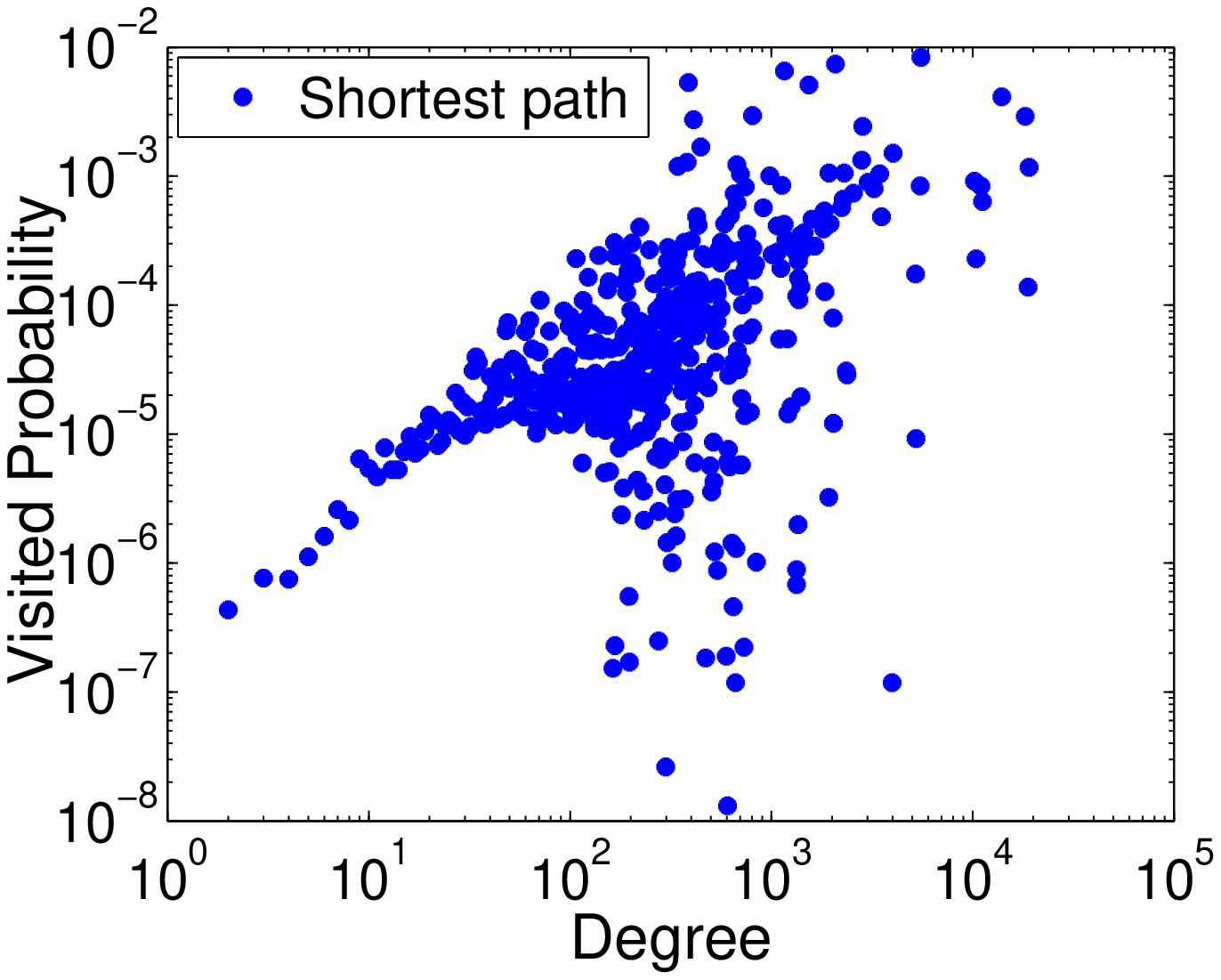}
}
\vspace{-1.7em}
\caption{The node degrees visited probabilities distributions of the real traces and the corresponding unweighted shortest paths on skitter dataset.}
\label{figure4}
\vspace{-0.5em}
\end{figure}

\subsubsection{Node Degree Visited Probability Distribution}
The node degrees visited probabilities distributions of the real
traces and the unweighted shortest paths of the two
datasets are shown in Figures \ref{figure3} and \ref{figure4},
respectively. It is interesting to note that in Figures
\ref{figure3:subfig:a} and \ref{figure4:subfig:a}, for some nodes
with high degrees, they might be rarely visited in the real routes,
which in another way implies that the assumption \textit{NDM} has made is unsatisfactory. It is also worthy noting that for some high degree nodes in unweighted shortest paths, their visiting possibilities are as large as $10^{-2}$ (a magnitude larger than the real traces with probabilities being $10^{-3}$), which indicates that \textit{USPM} is prone to visit high degree nodes more. In
addition, Figures \ref{figure3} and \ref{figure4}
show that when the node degrees are large, their visited probabilities observed from unweighted shortest paths, from $10^{-8}$
to $10^{-2}$, fluctuate more severely than those observed from the
real traces, from $10^{-5.5}$ to $10^{-3}$. Hence, \textit{USPM} differs significantly from the real case.

The above two phenomena motivate us to put forward better simulating models to more accurately characterize the two kinds of distributions above, especially the routes length distribution. 

\subsection{Local Information Based Model (\textit{LIM})}
In scale-free networks with degree distributions following
power-law, previous researches \cite{TrafficDynamicLocal,EfficientRouting}
have pointed out that unweighted shortest path based routing
strategy, that is \textit{USPM}, would reduce the communication
efficiency of networks. The reason is that such kind of strategy inclines to transmit traffic to the hubs with large degrees, which would easily introduce traffic jam to these hubs thereby inducing the decrease of the network communication capability. The situation would be severer if using \textit{NDM} since it always routes packets to the highest degree neighbors of nodes. To avoid this issue and so forth to better simulate the Internet routing, we propose a new model that uses the local information
\begin{equation}
\label{eq:lim}
W_{s\rightarrow i}=\frac{k_{i}^{\alpha}}{\sum_{j}k_{j}^{\alpha}}
\end{equation}
to assign weight to the edge between nodes $s$ and $i$ with $s$ the current node and $i$ the next routing node, where the sum runs over the neighbors of $s$, $k_i$ is the degree of node $i$, and $\alpha$ is a tunable parameter. We regard the graphs as bidirectional ones and
the weights of the bidirectional edges between two nodes are different, which accords to the different path features of up and down links in reality. After weighing, we carry out weighted shortest path method to simulate the real routes. In this model,
we could intuitively learn that here $\alpha<0$ means the links connected to the neighbors with
higher degrees might be preferentially introduced into the shortest
path, while $\alpha>0$, the situation would be reversed to choose the links connected to the lower degree neighbors with higher priority.

\subsection{Path Feature Based Model (\textit{PFM})}
In this subsection, we describe \textit{PFM} which uses the path features to add weights to the corresponding edges and then simulates the real Internet routes by conducting weighted shortest path algorithm. Also, the weights of the bidirectional edges between each pair of connected nodes are different. Weights of links ($x$) are assigned based on the bounded Pareto distribution, which is the simplest heavy-tailed distribution with the probability density function
\begin{equation}
\label{eq:pareto_distr}
p(x)=\frac{\alpha L^{\alpha} x^{-\alpha -
1}}{1 - (\frac{L}{M})^{\alpha}}, L > 0, M > L, \alpha > 0,
\end{equation}
where $\alpha$ determines the shape, $L$ denotes the minimal value, and $M$ denotes the maximal value.

This model is mainly driven by the fact that
in Internet heavy-tailed distributions have been observed in the context of traffic characterization, known as the self-similar nature of traffic \cite{WideAreaTraffic,WWWTraffic,ParetoModulatedModel}.
In addition, \cite{EndToEndRouting} observed high variability in path features, such as round-trip time, loss rate, and bandwidth, and wide variability is one of the characteristics of heavy-tailed distributions. Consequently, in this model, we allocate weights to edges according to the Pareto distribution of the corresponding path attributes, such as delay, loss rate, available bandwidth, and traffic load, and then simulate the real routes by performing weighted shortest path method. It is consistent with the real-world scenario that a packet is routed to the link with smaller delay, smaller loss rate, lighter traffic load, larger available bandwidth, and so on, in order to reach the target more efficiently and with a higher delivery success rate.

In summary, we present two new models with tunable parameters $\alpha$, which makes our models much more flexible than other models. In fact, different values of the parameter would generate different weights for the edges and stand for diverse physical link situations in real Internet routes. In the next section, we evaluate these models and determine the best configuration of our models.

\section{Evaluation and Discussion}
\label{sec:evaluation}
This section first evaluates our models, i.e., compares
the performance of \textit{LIM}, \textit{PFM}, \textit{NDM}
\cite{SimulatingInternetRoute}, and \textit{USPM} on the estimated accuracy of the structural properties of the real
network datasets. Then, we explain why our models are better than other existing models and discuss the implications of optimal parameters.

\subsection{Evaluation}
In this subsection, we evaluate our models on the two real datasets: iPlane and skitter. The evaluation examines their accuracies of simulating the routes length distribution and estimating the other network topological properties, including {\it average degree}, {\it degree distribution power-law exponent}, {\it clustering coefficient}, and {\it heterogeneity}. The simulation experiments of all the four models are conducted from 140 sources to common 20,419 destinations on iPlane dataset and from 21 sources to a common set of 157,290 destinations on skitter dataset as described in Section~\ref{sec:preliminaries}. For parameter configuration, we set $-5 \le \alpha \le 3$ for \textit{LIM} and we configure $L=10$, $M=|V|$, and $0 <
\alpha \le3$ for \textit{PFM}. We adopt the mean of 100
independent experiments as the final result for each $\alpha$.
\subsubsection{Routes Length Distribution}
\begin{figure*}[!t]
\centering
\subfigure[iPlane]{
\label{figure7:subfig:a}
\includegraphics[scale=0.265]{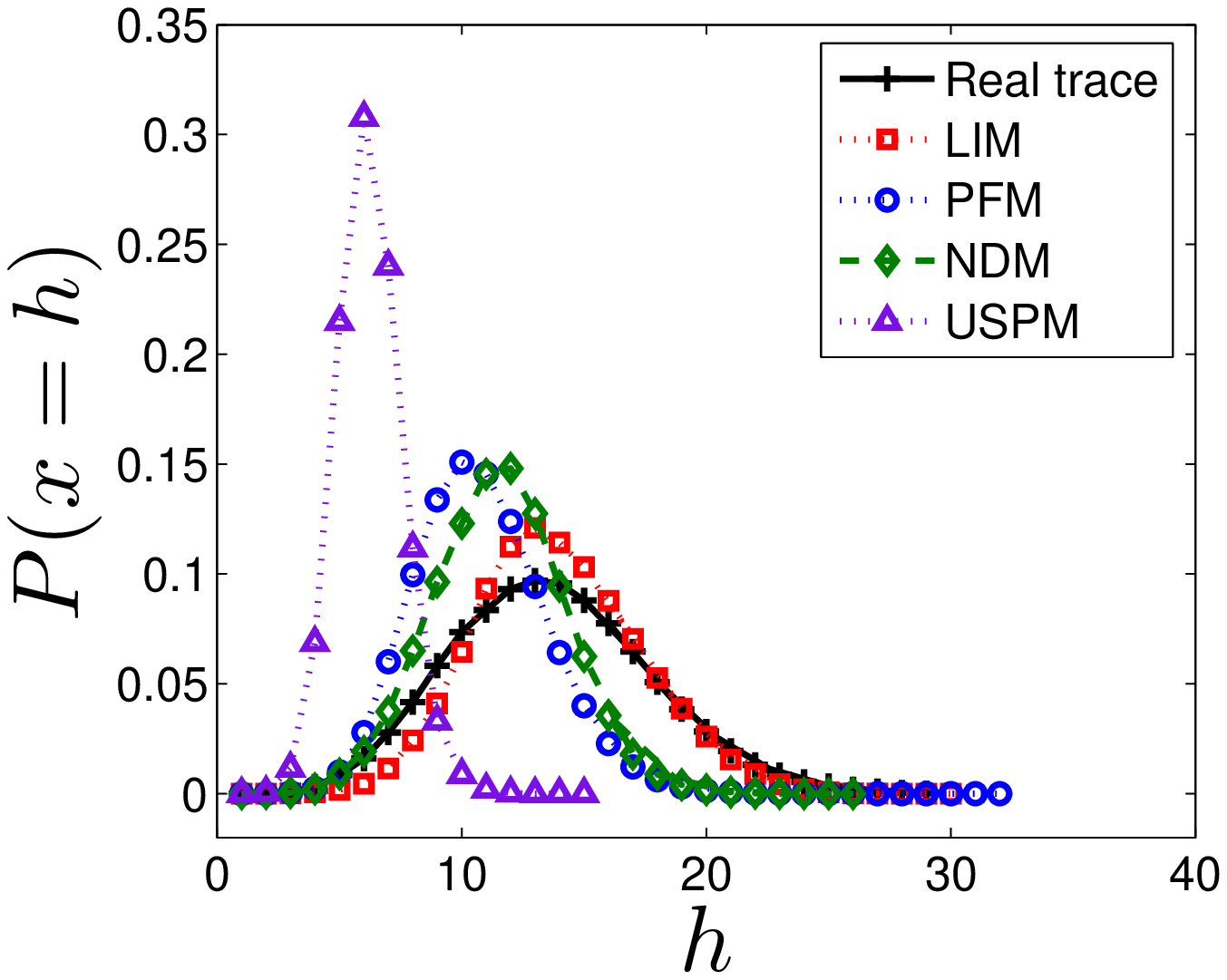}
}
\hspace{2em}
\subfigure[skitter]{
\label{figure7:subfig:b}
\includegraphics[scale=0.265]{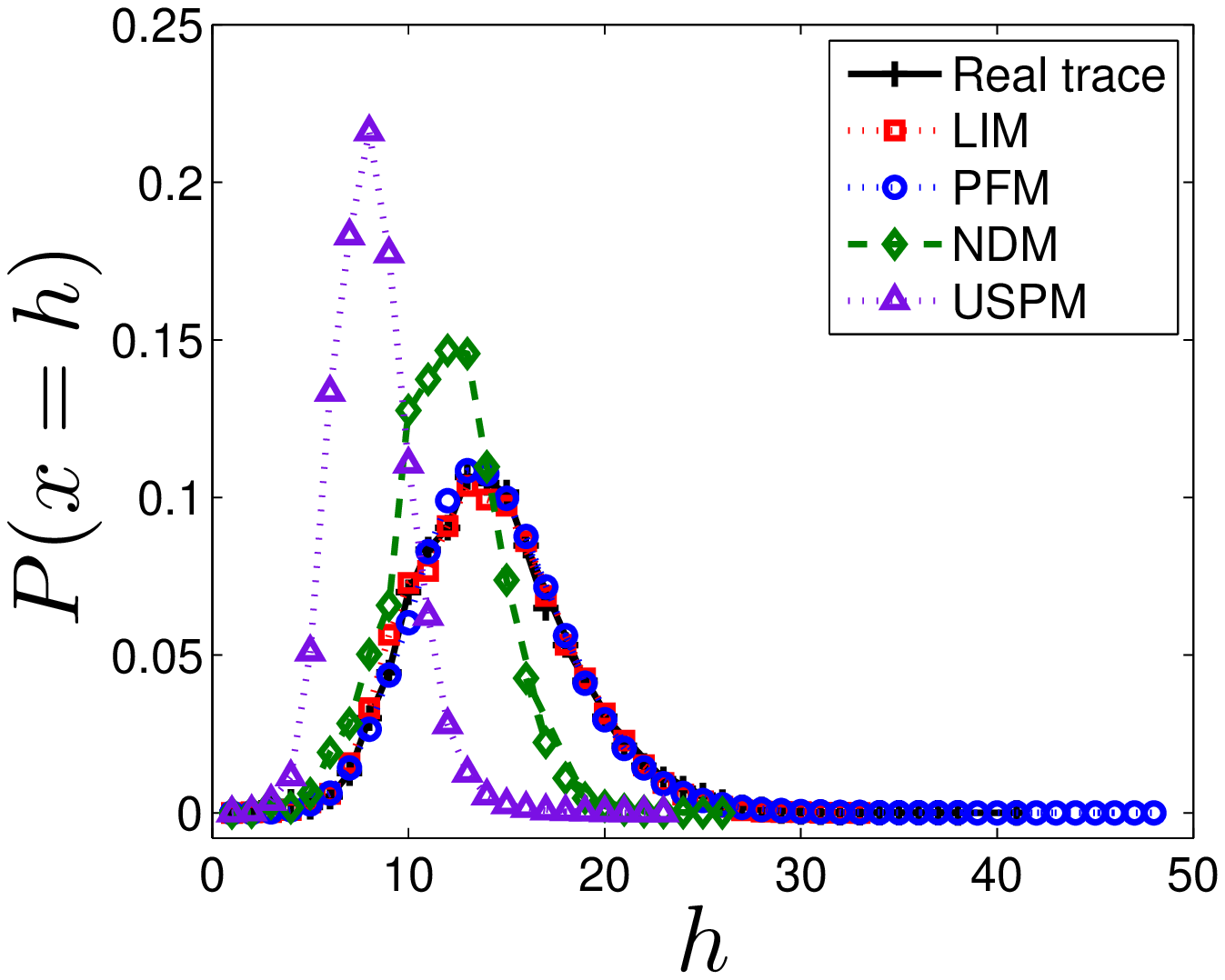}
}
\vspace{-1.7em}
\caption{The routes length distribution of the real
traces and the four models on two real datasets: iPlane and skitter, respectively.
Here the hop of routes is denoted as $h$.}
\label{figure7}
\vspace{-0.5em}
\end{figure*}

Figure~\ref{figure7} presents the routes length distribution of the real
traces and the four models on the two real datasets, respectively. The routes length distributions of \textit{LIM} and \textit{PFM} are their optimal ones obtained by tuning the parameter $\alpha$. On iPlane, the best $\alpha$ for \textit{LIM} and
\textit{PFM} is $\alpha=1.6$ and $\alpha=0.1$, respectively; when on skitter dataset, $\alpha=0.9$ and $\alpha=0.2$ is the best for \textit{LIM} and \textit{PFM}, respectively. It can be seen that the routes length
distributions of \textit{LIM} on iPlane and skitter datasets are
almost the same with those of the real trace. While, with regard to the other
three models, there is still a considerable gap as compared to the real one,
especially for \textit{NDM} and \textit{USPM}. Consequently, it
can be concluded that \textit{LIM} is the best model to
simulate the routes length distribution of the real routes. Note that we will not consider \textit{USPM} in the following evaluation for its extremely bad simulation of the routes length distribution of the real Internet routes.

\begin{figure*}[!t]
\centering
\subfigure[iPlane]{
\label{figure8:subfig:a}
\includegraphics[scale=0.265]{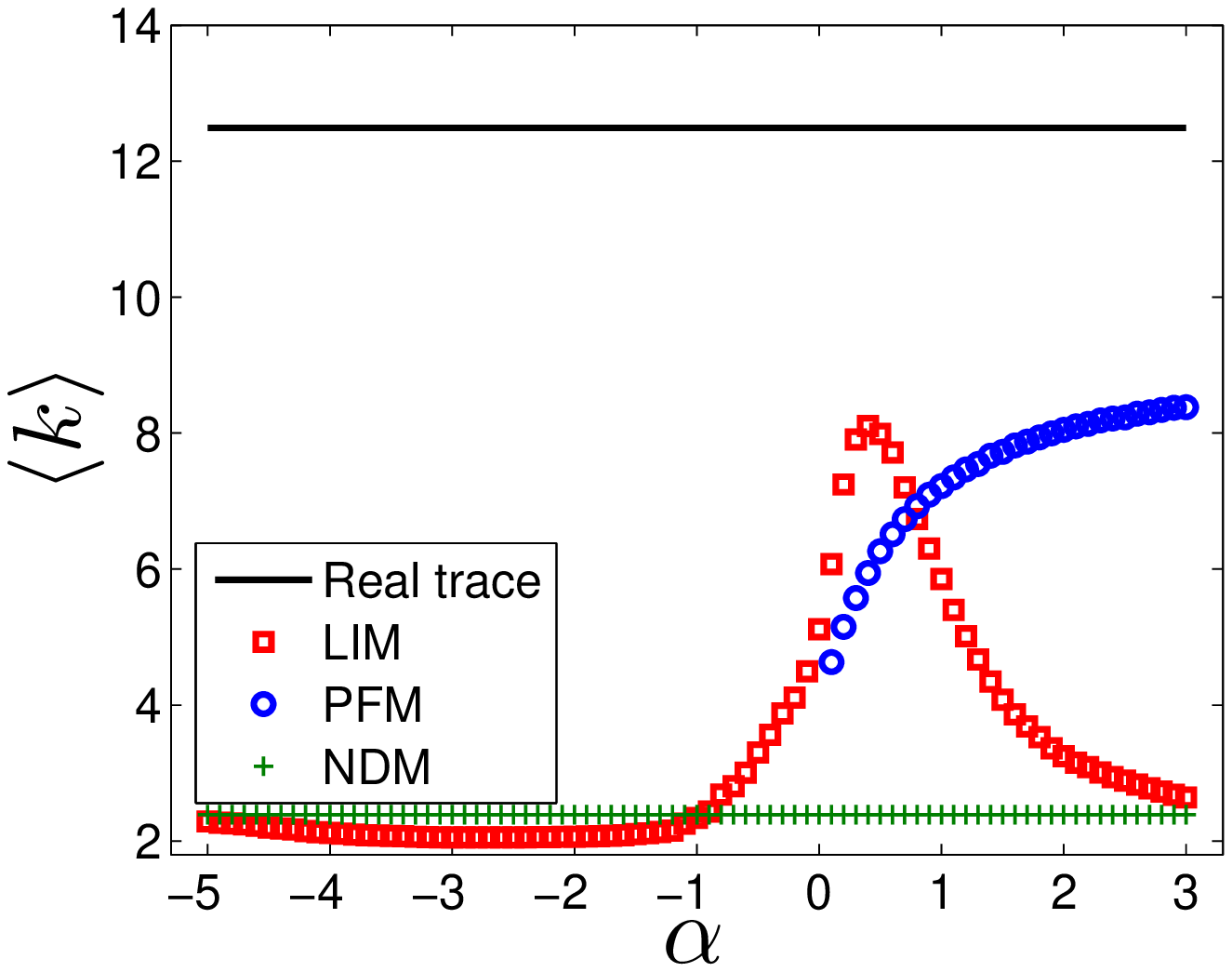}
}
\hspace{2em}
\subfigure[skitter]{
\label{figure8:subfig:b}
\includegraphics[scale=0.265]{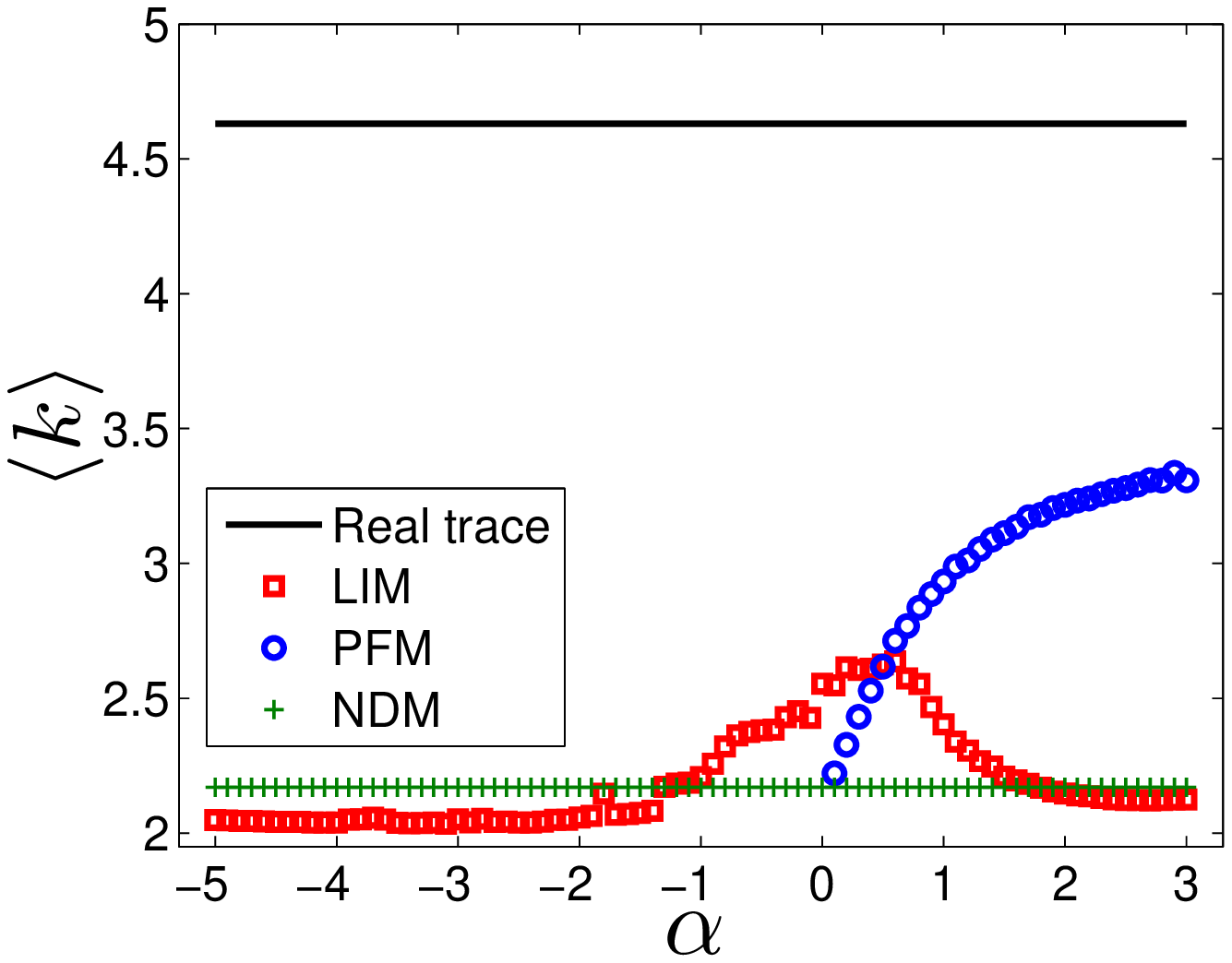}
}
\vspace{-1.7em}
\caption{The {\it average degree} of the sampled graphs by using the three models, compared with that of $G$, on two real datasets: iPlane and skitter, respectively.}
\label{figure8}
\vspace{-0.8em}
\end{figure*}

\subsubsection{Average Degree}
Fig. \ref{figure8} shows the {\it average degree} of $G$ and
those of the sampled graphs by adopting the three models:
\textit{LIM}, \textit{PFM}, and \textit{NDM}. It indicates that our
two models, in most cases, are much better than \textit{NDM}. \textit{LIM} reaches its peaks on the two datasets when $\alpha = 0.4$ and $\alpha = 0.6$, respectively; \textit{PFM} reaches its peaks on both datasets when
$\alpha=3$. Besides,
\textit{PFM} is always better than \textit{NDM}; \textit{LIM} is much better than
\textit{NDM} on the two datasets when $\alpha >
-1$ and $-1.4 < \alpha < 1.9$, respectively. Hence, our two models are more effective
than \textit{NDM} on sampling the {\it average degree} of the underlying network.

\begin{figure*}[!t]
\centering
\subfigure[iPlane]{
\label{figure9:subfig:a}
\includegraphics[scale=0.265]{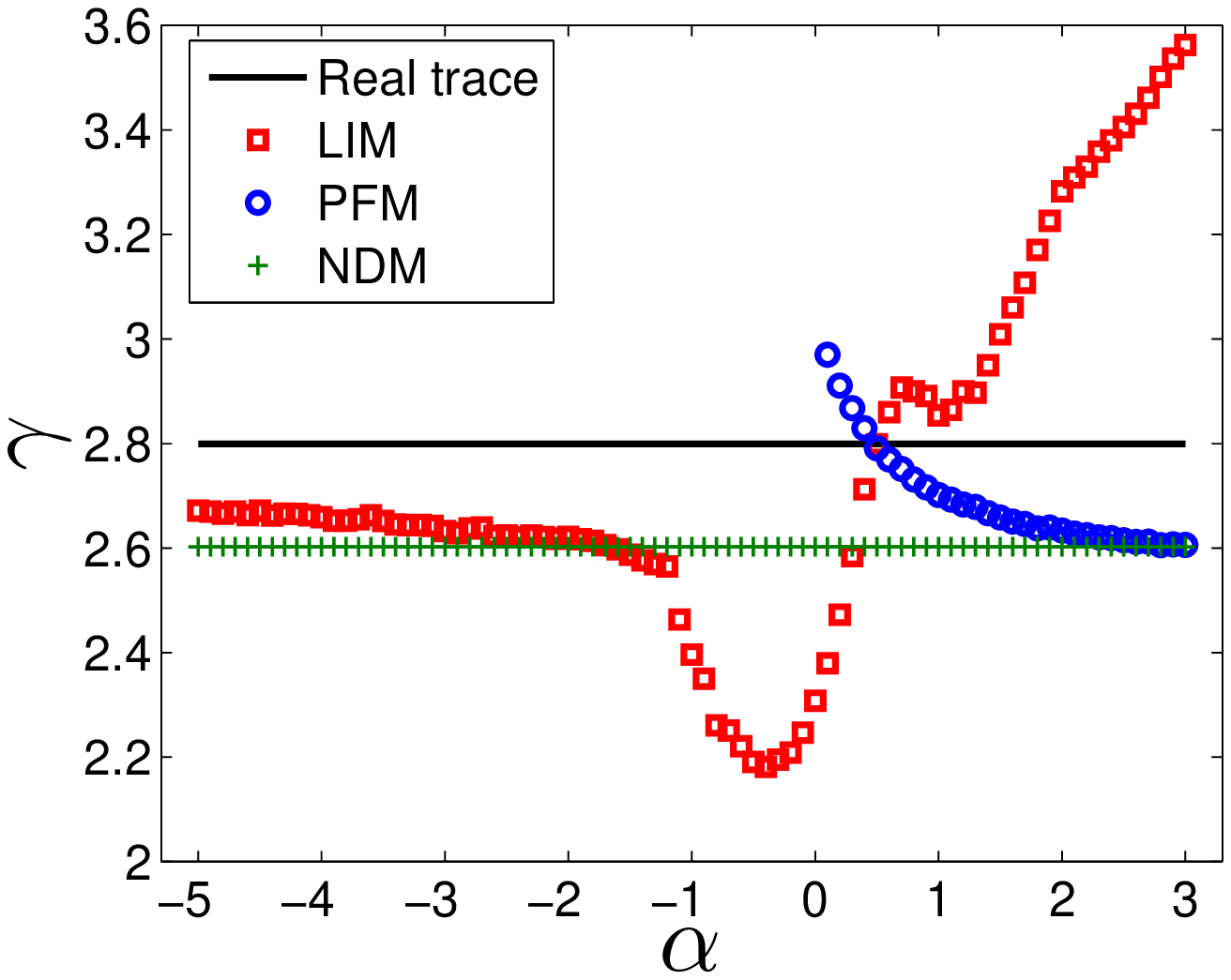}
}
\hspace{2em}
\subfigure[skitter]{
\label{figure9:subfig:b}
\includegraphics[scale=0.265]{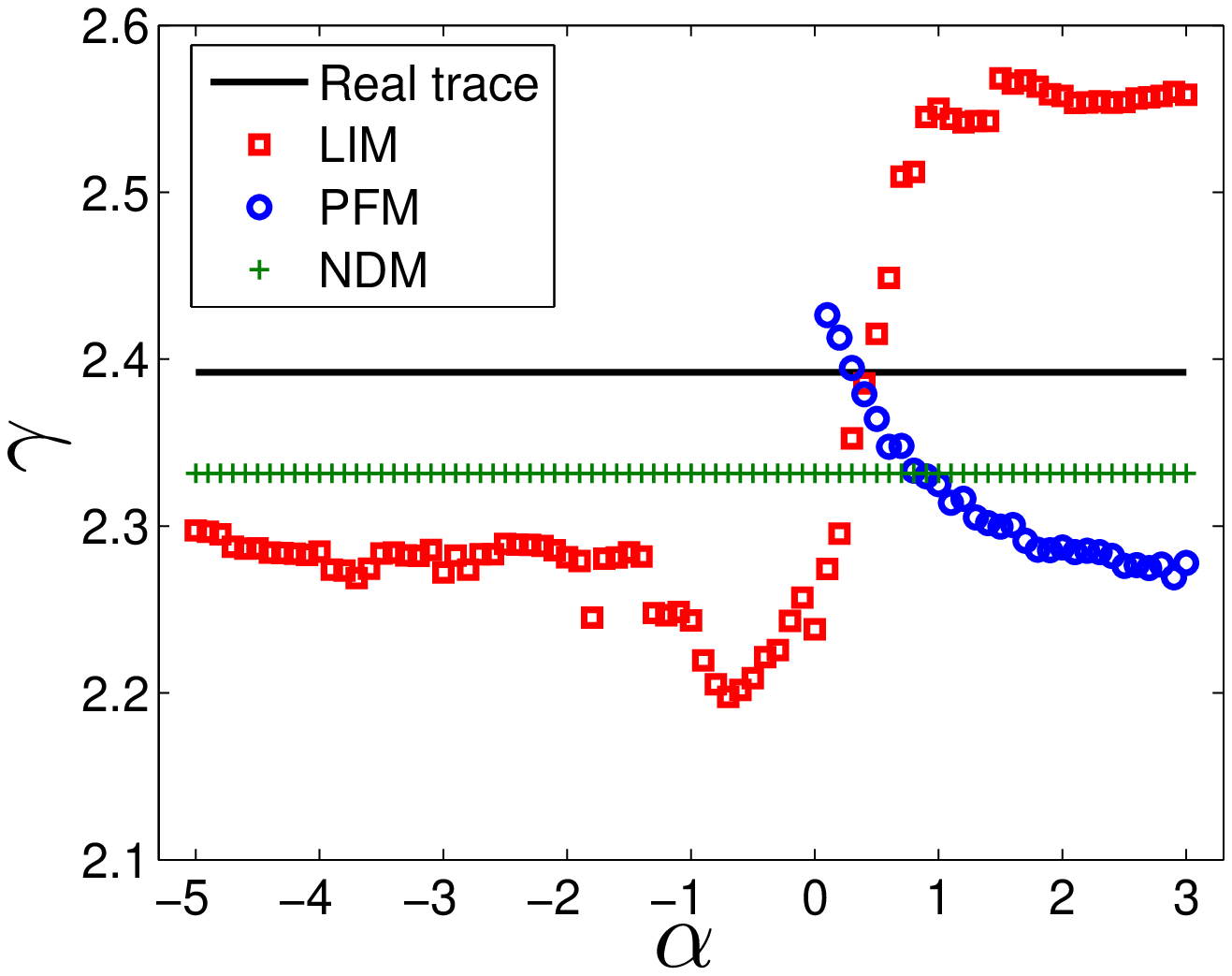}
}
\vspace{-1.7em}
\caption{The {\it degree distribution} power-law exponent $\gamma$ of the sampled graphs by using the three models, compared with that of $G$, on the two real datasets, respectively.}
\label{figure9}
\vspace{-0.5em}
\end{figure*}

\subsubsection{Power-law Degree Exponent}
Fig. \ref{figure9} presents the {\it degree distribution} power-law exponent $\gamma$, which is calculated with the
least square method \cite{LeastSquareMethod}, of $G$ and those of
the sampled graphs by adopting the three models. It can be observed that our models
could achieve the same values with the real trace.
\textit{LIM} obtains the same values with $G$ on both iPlane and skitter datasets when $\alpha =
0.5$; \textit{PFM} gets the same values with $G$ on
the two datasets when $\alpha=0.5$ and
$\alpha=0.3$, respectively. Moreover, \textit{PFM} is, in most cases, better
than \textit{NDM}; \textit{LIM} is better than \textit{NDM} on the two network datasets when $0.2 < \alpha < 1.7$ and $0.2 < \alpha <
1.6$, respectively. Consequently, we can infer that our two models perform much better than \textit{NDM} on simulating the {\it degree distribution}.

\begin{figure*}[!t]
\centering
\subfigure[iPlane]{
\label{figure10:subfig:a}
\includegraphics[scale=0.265]{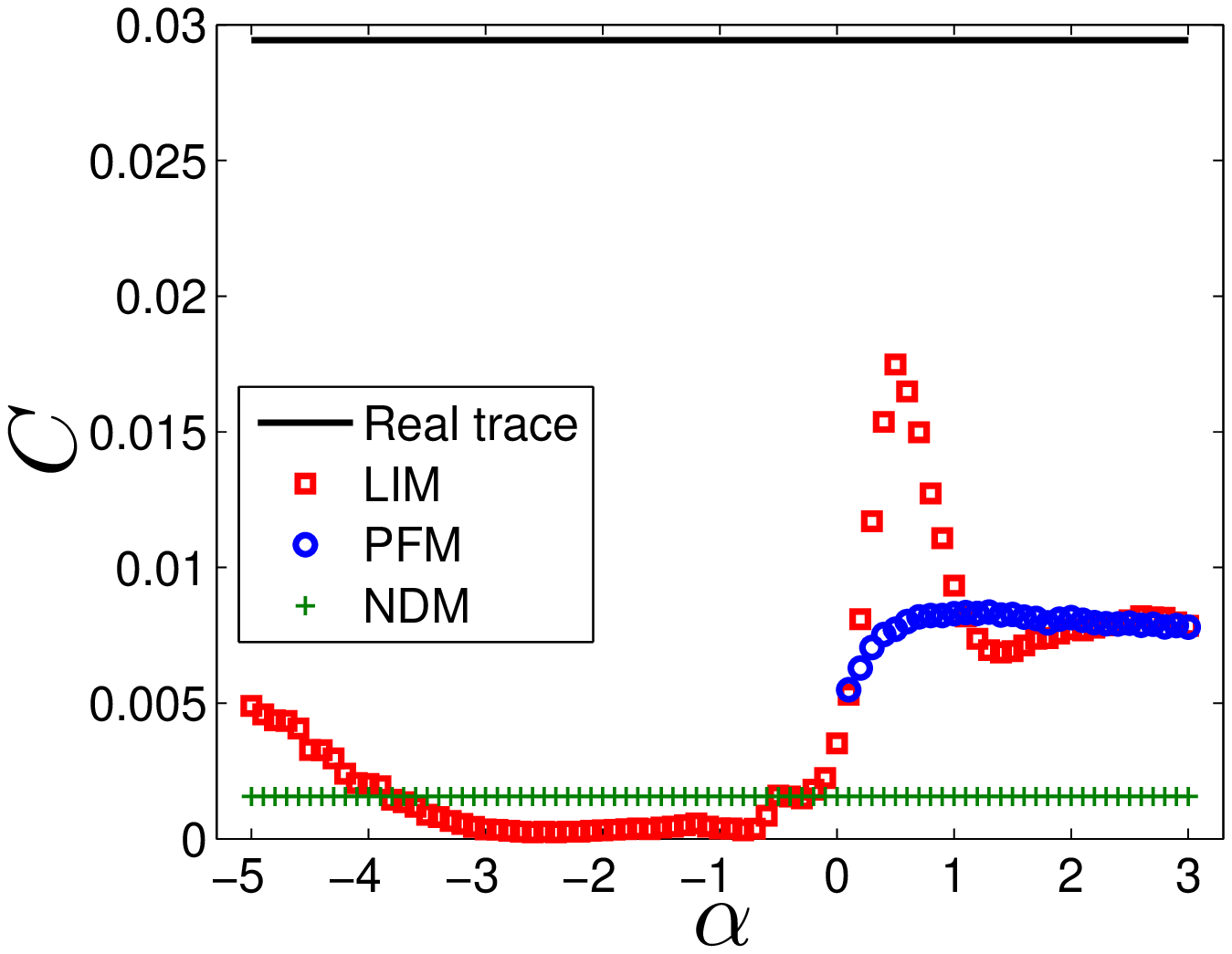}
}
\hspace{2em}
\subfigure[skitter]{
\label{figure10:subfig:b}
\includegraphics[scale=0.265]{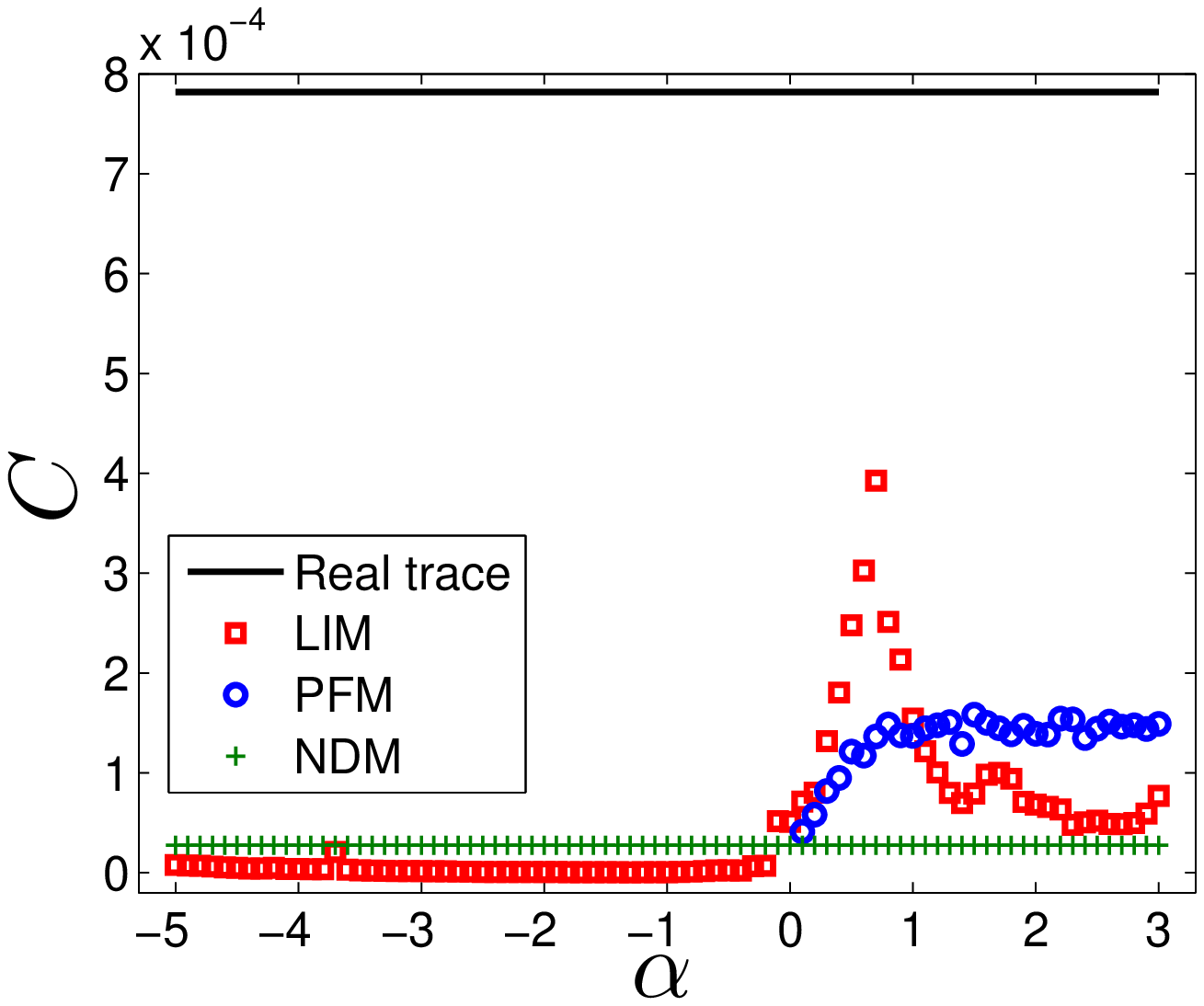}
}
\vspace{-1.7em}
\caption{The {\it clustering coefficient} of the sampled graphs by using the three models, compared with that of $G$, on two real datasets: iPlane and skitter, respectively.}
\label{figure10}
\vspace{-0.8em}
\end{figure*}

\subsubsection{Clustering Coefficient}
Fig. \ref{figure10} shows the {\it clustering coefficient} of $G$ and those of the sampled graphs by adopting the three models. It indicates that our two models, in most cases, are much better than \textit{NDM}. \textit{LIM} reaches its peaks on the two datasets, iPlane and skitter, when $\alpha = 0.5$ and $\alpha = 0.7$, respectively; \textit{PFM} reaches its peaks on both datasets when $\alpha=3$. Further, \textit{PFM} is always better than \textit{NDM} and \textit{LIM} is much more effective than \textit{NDM} on both datasets when $\alpha > 0$. Therefore, our two models could obtain a more accurate estimation of the {\it clustering coefficient} of the underlying network.

\begin{figure*}[!t]
\centering
\subfigure[iPlane]{
\label{figure11:subfig:a}
\includegraphics[scale=0.265]{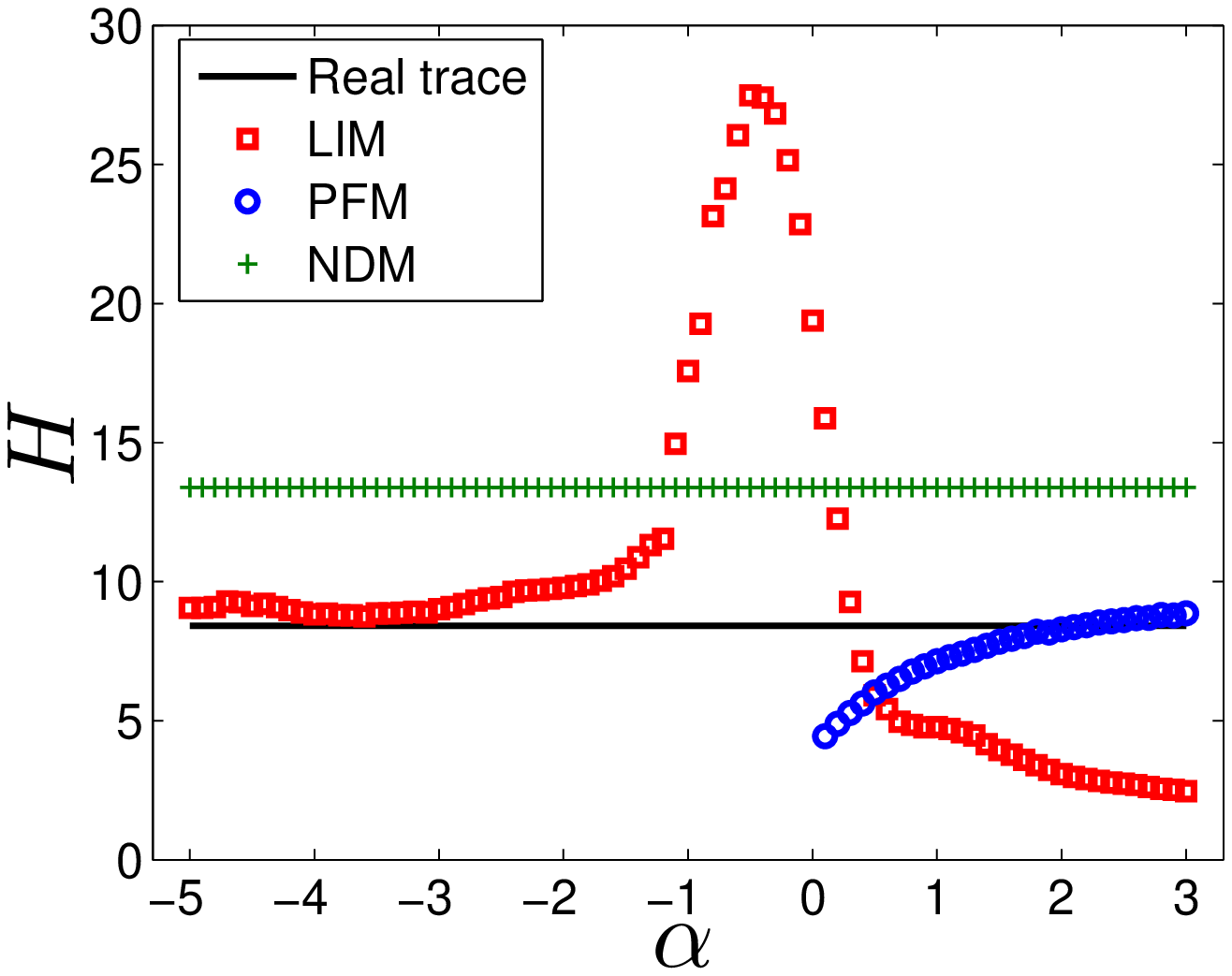}
}
\hspace{2em}
\subfigure[skitter]{
\label{figure11:subfig:b}
\includegraphics[scale=0.265]{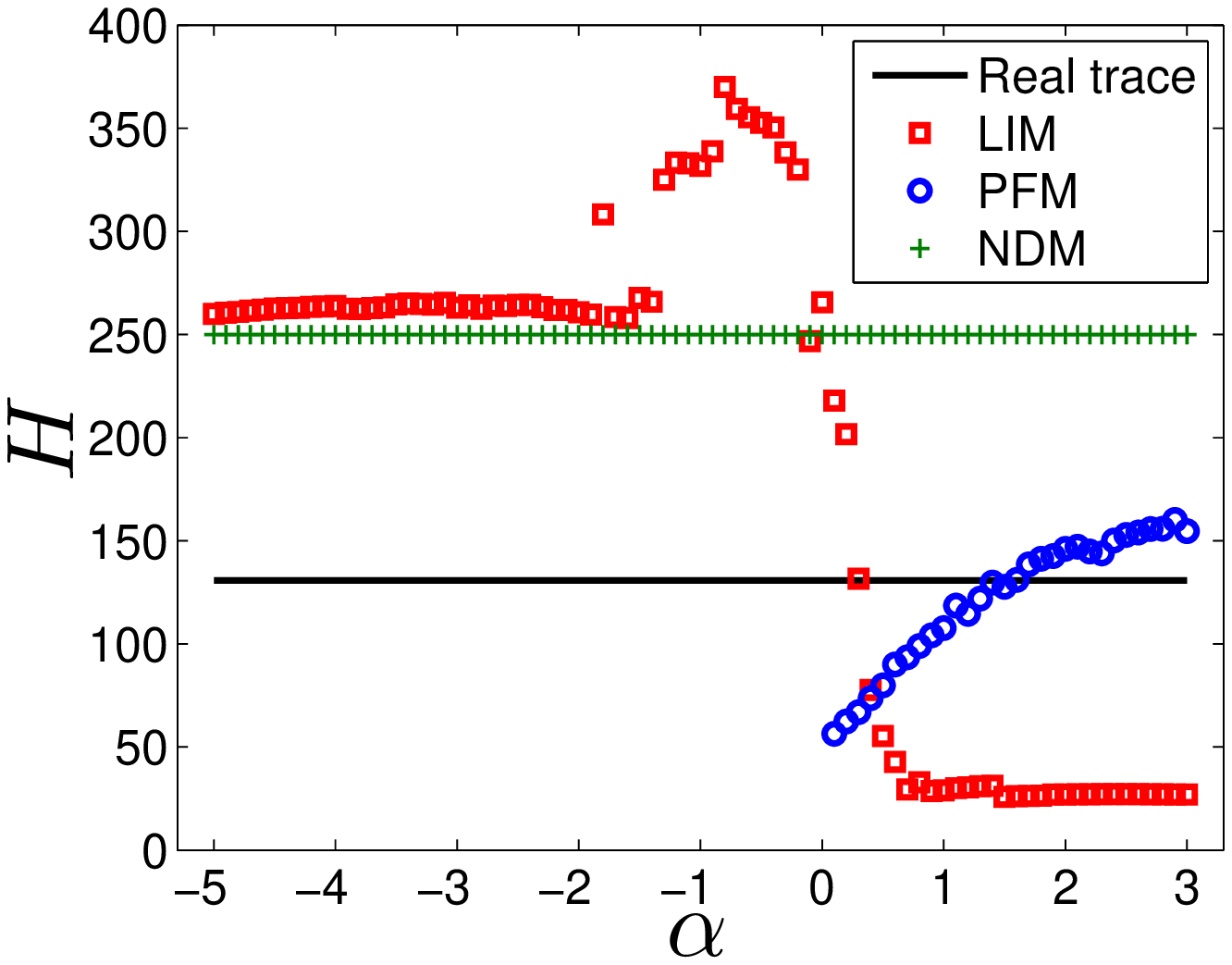}
}
\vspace{-1.7em}
\caption{The {\it heterogeneity} of the sampled graphs by using the three models, compared with that of $G$, on two real datasets: iPlane and skitter, respectively.}
\label{figure11}
\vspace{-0.5em}
\end{figure*}

\subsubsection{Heterogeneity}
Fig. \ref{figure11} presents the {\it heterogeneity} of $G$
and those of the sampled graphs by adopting the three models. It can be found that
our models could achieve almost the same values with the real trace. \textit{LIM} obtains the same values with $G$ on both datasets when $\alpha = 0.4$; \textit{PFM} gets the same values with $G$ on iPlane and skitter datasets when $\alpha=2.2$ and $\alpha=1.6$, respectively. Besides, \textit{PFM} is always better than \textit{NDM}; \textit{LIM} is better
than \textit{NDM} on iPlane dataset when $\alpha < -1$ and $\alpha > 0$, and it is better than \textit{NDM} on
skitter dataset when $\alpha > 0$. Consequently, it could be concluded that
our models could sample the {\it heterogeneity} of the network with a convincing fidelity.

\subsection{Evaluation Conclusion}
In previous subsection, we evaluate our models both on routes length
distribution and on the other network topological properties, such as {\it
average degree}, {\it degree distribution}, {\it clustering
coefficient}, and {\it heterogeneity}, on the two real datasets: iPlane
and skitter. The evaluation implies that our two models are more effective than \textit{NDM} not only on the simulating of routes length distribution, but also on the sampling of the other structural properties. In addition, our models
do not give rise to more consuming time and have the similar
computing complexity as \textit{NDM} and \textit{USPM}. However,
\textit{PFM} is much worse than \textit{LIM} on modeling the routes
length distribution. Particularly, it has been found in reality that for self-similar traffic,
$\alpha$ for \textit{PFM} should be in the region of $(1,2)$
\cite{ParetoModulatedModel}, which is not overlapped with the proper
range we find above. Therefore, for a better simulation of the real
Internet routes, we recommend \textit{LIM} instead of \textit{PFM}.
But it is also important to point out that one has to
choose different proper $\alpha$ for \textit{LIM} in order to
achieve the best sampling of different network properties, which is shown in
Table \ref{Different alpha for different properties}. However,
it is interesting to observe from Table \ref{Different alpha for
different properties} and also from previous subsection that when
setting $0 < \alpha < 2$, almost all of the topological properties could be under
accurate estimation. Thus, to better model the Internet routing, \textit{LIM} with $0 < \alpha < 2$ is recommended; in
addition, researchers could set $\alpha$ to be around 0.5 to more
accurately estimate the other network structural properties except the routes
length distribution.
\begin{table}[!t]
\footnotesize
\caption{The best value of \textit{LIM}($\alpha$) for simulating and sampling different network topological properties of the real Internet, including iPlane and skitter datasets.}
\label{Different alpha for different properties}
\vspace{-1em}
\centering
\setlength{\tabcolsep}{4pt}
\begin{tabular}{c|c||c}
\hline \hline
    Dataset & Topological property & \textit{LIM}($\alpha$) \\
\hline
    & Routes length distribution & 1.6  \\
    \cline{2-3}
           & $\langle k \rangle$ & 0.4  \\
    \cline{2-3}
    iPlane & $\gamma$ & 0.5  \\
    \cline{2-3}
           & $C$ & 0.5  \\
    \cline{2-3}
           & $H$ & 0.4  \\
\hline
    & Routes length distribution & 0.9  \\
    \cline{2-3}
           & $\langle k \rangle$ & 0.6  \\
    \cline{2-3}
    skitter & $\gamma$ & 0.5  \\
    \cline{2-3}
           & $C$ & 0.7  \\
    \cline{2-3}
           & $H$ & 0.4  \\
\hline \hline
\end{tabular}
\end{table}

\subsection{Discussion}
In this subsection, we explain why the proper value of
$\alpha$ for \textit{LIM} should be in the range of $(0,2)$ and the implications of such range.

\begin{figure*}[!t]
\centering
\subfigure[iPlane]{
\label{figure12:subfig:a}
\includegraphics[scale=0.3]{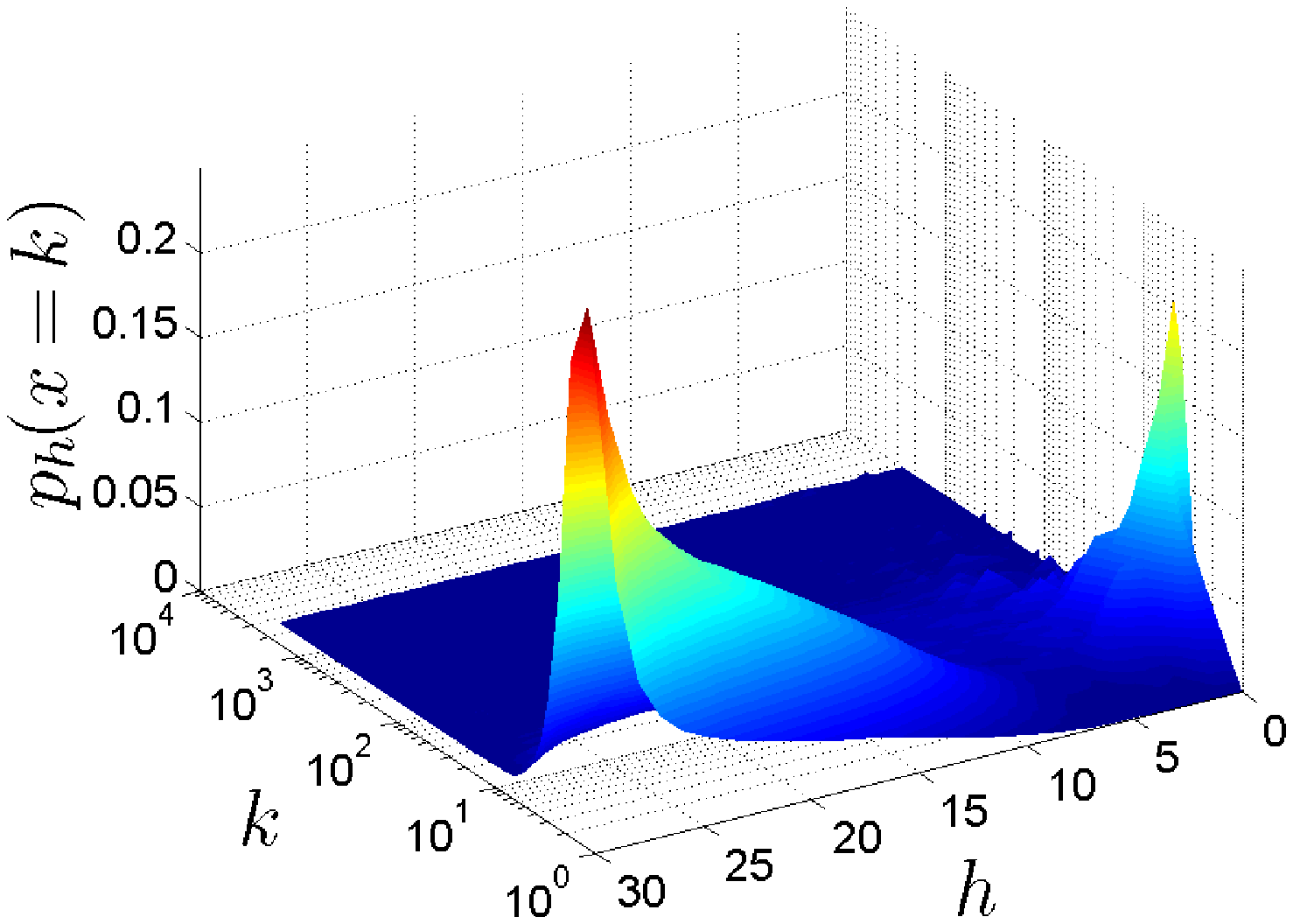}
}
\hspace{2em}
\subfigure[skitter]{
\label{figure12:subfig:b}
\includegraphics[scale=0.3]{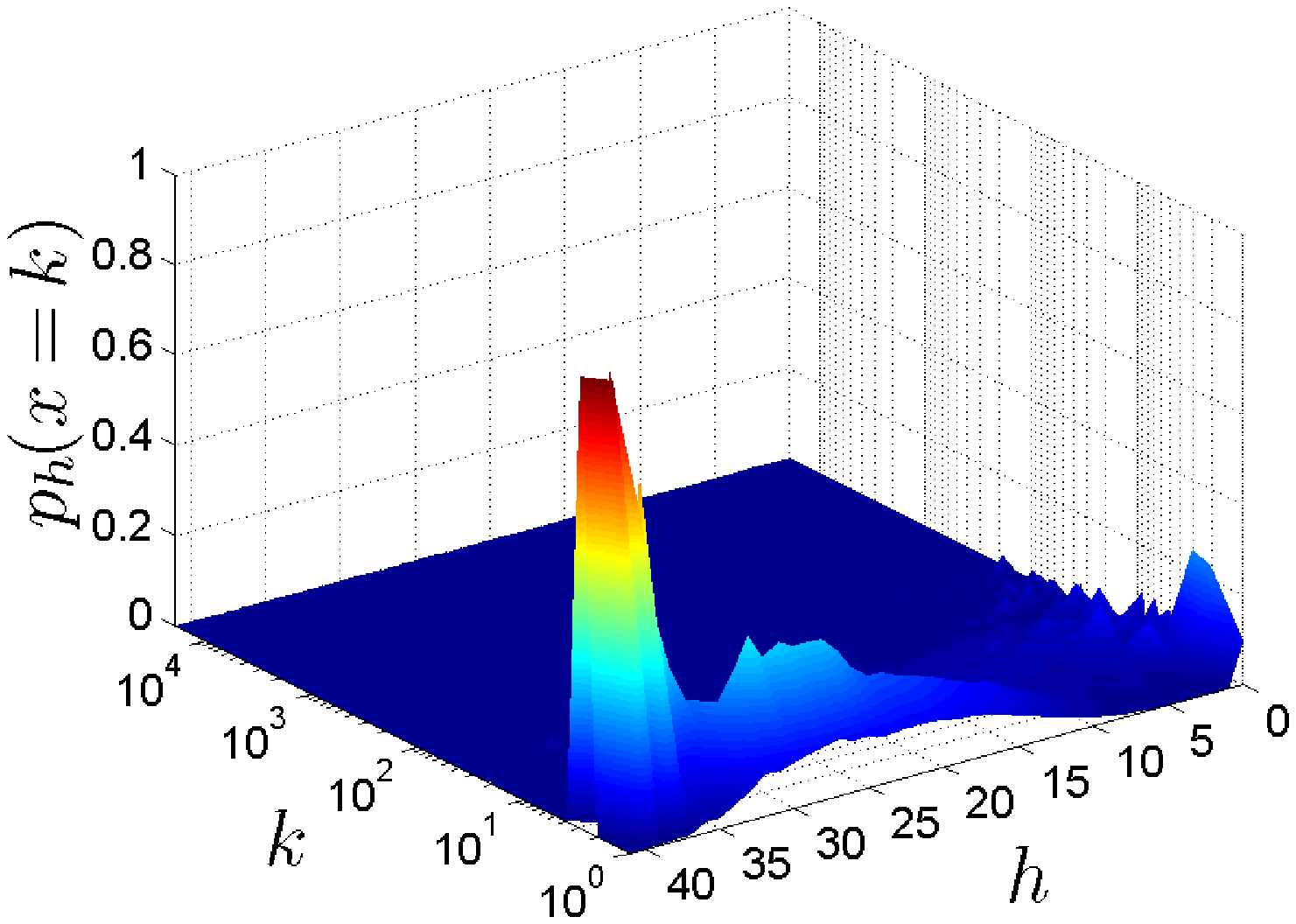}
}
\vspace{-1.7em}
\caption{The degree distribution of routes in each hop extracted from the real traceroute traces of iPlane and skitter datasets, respectively. ($k$ is node degree; $h$ represents hop; $p_h(x=k)$ represents the possibility of $k$ in the node degree distribution of $h$.)}
\label{figure12}
\vspace{-0.8em}
\end{figure*}

As stated in
\cite{TrafficDynamicLocal,EfficientRouting}
that when $\alpha$ in Eq.~\ref{eq:lim} is positive, the traffic
load on each node and also the packet traveling time would follow power-law
distributions, which indicates the highly heterogeneous status; when
$\alpha < 0$, the traffic load on each node and the packet traveling
time would display as Poisson distribution and exponential
distribution, respectively, which represents a homogeneous state.
Additionally, a positive $\alpha$ in \cite{TrafficDynamicLocal,EfficientRouting} represents that packets tend to be routed
to the nodes with large degrees, which induces the hubs easily turn
to be jammed and then decreases the communication capability of the
networks. While, a negative $\alpha$ in \cite{TrafficDynamicLocal,EfficientRouting} illustrates that the
communication networks, in some cases, try to shunt some traffic
from the hubs to the nodes with small degrees in order to reduce the
loads of those hubs. Therefore, in those research work, a negative $\alpha$ is better than a
positive $\alpha$. While the negative $\alpha$ in \cite{TrafficDynamicLocal,EfficientRouting} is corresponding to the positive $\alpha$ in our research because we adopt the values of Eq.~\ref{eq:lim}
as the weights of the edges and then perform the weighted shortest
path algorithm. There is also an interesting phenomenon that
in the real Internet routes, packets incline to move to nodes
with small degrees, which could be seen from Fig. \ref{figure12}
that presents the degree distribution of routes in each hop
extracted from the real traceroute traces of iPlane and skitter
datasets, respectively. It shows that, in most cases, routes with their routing at each hop tend to move
to the small degree nodes, especially the nodes with degrees from 1 to 10.
This phenomenon surprisingly justifies that the route strategy with $\alpha
< 0$ in \cite{TrafficDynamicLocal,EfficientRouting}
and \textit{LIM} with $\alpha > 0$ in our work reflect the realistic
situation in the Internet routes. Meanwhile, it could be used to
explain the poor performance of \textit{USPM} and \textit{NDM}, which are
their intents of targeting the large degree nodes.

\begin{figure}[!t]
\centering
\includegraphics[scale=0.25]{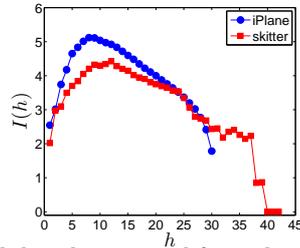}
\vspace{-1.7em}
\caption{The entropy of each hop $h$ extracted from the real traceroute traces on iPlane and skitter datasets, respectively.}
\label{figure14}
\vspace{-0.5em}
\end{figure}

Moreover, \cite{TrafficDynamicLocal,EfficientRouting}
declared that the packet traveling time increases as the decrease of the negative $\alpha$ in their route
strategy. It is the cost that the network should pay for reducing the burdens of hubs by transferring some traffic to low degree nodes, which, in purpose, enhances the communication capability of the network. This fact also reflects the phenomenon that the increase of the average length of the routes as the increase of the positive $\alpha$ in \textit{LIM}.
However, it is also stated in \cite{TrafficDynamicLocal,EfficientRouting} that $|\alpha|$ is far from the larger the better, because too large $|\alpha|$ may sharply increase the packet traveling time, which would in turn decrease the network's communication capability too. In addition, as shown in Fig. \ref{figure12}, especially in Fig. \ref{figure12:subfig:b}, the node degree (from 1 to 100) distributes relatively evenly for some small hops (from 5 to 15), which indicates that the low degree nodes are relatively less dominant in these hops as compared to other large ones. Moreover, we define the entropy of each hop as
\begin{equation}
\label{eq:entropy}
I(h)=-\sum_{k}{p_h(k)\log{p_h(k)}},
\vspace{-0.5em}
\end{equation}
where $h$ is the hop, $k$ is the node degree that appears in this hop, and $p_h(k)$ is its probability. It is worthy noting that here we omit the $k$ with $p_h(k)=0$. According to the definition, larger $I(h)$ means in that hop $h$, each node degree distributes relatively evenly, while lower $I(h)$ means some node degrees are dominant in this hop. Fig. \ref{figure14} shows that on both iPlane and skitter datasets, when the hop is in the range of $(5,15)$, different node degrees distribute somewhat evenly, while in other hops, mainly the low node degrees dominate. It tells us that for many small hops, the nodes with intermediate degrees (from 10 to 100) would be considered in the routing, while too large $\alpha$ for \textit{LIM} would indeed ignore them entirely. Consequently, $\alpha$ in \textit{LIM} should be in the range of $(0,2)$ instead of exceeding 2.

To sum up, our simulating model \textit{LIM}, with $\alpha$ in the range of $(0,2)$, represents the real situation of the
Internet routes. It actually reflects the design philosophy of Internet, trying to achieve a trade-off between network communication efficiency (packet traveling time) and traffic load balance among nodes. The $\alpha$ range $(0,2)$ for \textit{LIM} would reach a win-win situation on both aspects. Thus, \textit{LIM} is a better model to simulate the routes in the Internet.

\section{Conclusion}
\label{sec:conclusion}

In this paper, we deeply study the real Internet routes and propose two novel models to well simulate the
Internet routing. Through thoroughly comparison with existing models, we
find that one of our models, \textit{LIM}, could outstandingly
simulate the routes length distribution and better estimate the other
topological properties of the Internet topology with proper
configurations of the parameter $\alpha$. Besides, we recommend research
community using \textit{LIM} with $\alpha$ in the range of $(0,2)$ to achieve better estimation on the
overall properties of the real routes in Internet, more specifically with $\alpha$ around 0.5 to more
accurately estimate the other network structural properties except the routes
length distribution.

\end{document}